# MEASUREMENT OF SOUND FIELDS USING MOVING MICROPHONES

*Fabrice Katzberg, Radoslaw Mazur, Marco Maass, Philipp Koch, and Alfred Mertins*

Institute for Signal Processing
University of Lübeck
Ratzeburger Allee 160, 23562 Lübeck, Germany

**ABSTRACT**

The sampling of sound fields involves the measurement of spatially dependent room impulse responses, where the Nyquist-Shannon sampling theorem applies in both the temporal and spatial domain. Therefore, sampling inside a volume of interest requires a huge number of sampling points in space, which comes along with further difficulties such as exact microphone positioning and calibration of multiple microphones. In this paper, we present a method for measuring sound fields using moving microphones whose trajectories are known to the algorithm. At that, the number of microphones is customizable by trading measurement effort against sampling time. Through spatial interpolation of the dynamic measurements, a system of linear equations is set up which allows for the reconstruction of the entire sound field inside the volume of interest.

*Index Terms—* Room impulse response, plenacoustic function, dynamic measurement, microphone array, perfect sequence

## 1. INTRODUCTION

Sound fields within closed rooms are characterized by reverberation, which decreases the performance of applications such as multichannel sound systems or speech recognition devices, since they assume a free-field environment. However, exact knowledge of the sound field inside a volume of interest enables us to remove acoustic scene effects by listening room compensation, and thus allows for a performance improvement within a wide range of applications.

There are several stationary methods available for the measurement of room impulse responses (RIRs). Common approaches are the use of perfect sequences [1, 2], maximum-length sequences (MLS) [3, 4], and exponential sine sweeps [5]. In all of these cases, the RIRs can be obtained by correlation techniques with a very low computational demand. For setups with time-varying RIRs, such as acoustic echo cancellation (AEC) [6], the common approach is to demand a minimum mean squared error (MMSE) between the measured signal and the output of an adaptive filter that is excited by the same input as the loudspeaker. Of course, the method can also be used to measure time-invariant RIRs. In this case, an excitation with white noise would be preferable, because it provides maximum convergence speed [7]. In [8, 9], the concept of the plenacoustic function (PAF) has been introduced. Generally, the PAF encapsulates the information on the entire set of spatio-temporal RIRs for any position in space. The sampling of the PAF by considering equidistant sampling points in space that satisfy the Nyquist-Shannon sampling theorem is not practical. In [10], a setup with moving microphones was studied. Applications include the simulation of time-varying channels that are governed by the wave equation and the computation of frequency-dependent update rules for adaptive algorithms. There is also an existing method for the dynamic measurement of a set of RIRs using one microphone that moves along a given trajectory [11]. Despite the motion of the microphone, the RIRs are reconstructed for all points along the trajectory. In particular, motion along a line and a circle have been studied. To make this principle work, a specially designed input signal is needed, and the speed of the microphone must be constant and is restricted to an upper limit. Quite different from RIR measurements with a fixed microphone, the excitation signal must not contain all audio frequencies, but only a certain subset. The omitted frequencies are essentially generated through the Doppler effect. The technique is suited to record relatively short RIRs and has also been proposed to record head related transfer functions (HRTFs) with a microphone that moves along a circular trajectory around the head.

In this paper, we propose an approach which allows for the measurement of the PAF using a moving array with a manageable number of microphones. Through spatial convolution, and interpolation of the dynamic measurements, a system of linear equations is set up which allows for a reconstruction of the sound field inside the considered volume. This method is totally different from the abovementioned dynamic measurement procedure. It does not require specific demands, neither on the excitation signal nor on the speed of the microphone or microphone array.

The paper uses the following notations. The operator $\text{diag}\{\cdot\}$ turns a vector into a diagonal matrix. The trace of a matrix is indicated by $\text{tr}\{\cdot\}$. $\boldsymbol{I}_{D\times D}$ denotes the $D \times D$ identity matrix. The unit pulse is represented by $\delta(n)$, which equals 1 for $n = 0$ and 0 elsewhere. The modulo operation is denoted by $a \bmod b$, which gives the remainder after the division of $a$ by $b$.

## 2. UNIFORM SAMPLING OF SOUND FIELDS

Measuring the sound field is basically a sampling problem. Under the assumption that the sound field is bandlimited, it can be reconstructed through equidistant sampling in both time and space dimensions. In the following, we first introduce the notation for the plenacoustic function and consider its uniform sampling in time domain. Then we describe its sampling in space.

### 2.1. The Plenacoustic Function

For a fixed pair of source and listener positions in space, the room impulse response $h(t)$ characterizes the time $t$ dependent received sound wave that results from a Dirac impulse emitted at time $t = 0$.

This work has been supported by the German Research Foundation under Grants No. ME 1170/8-1 and ME 1170/10-1.

For an excitation signal $s(t)$, the observed signal is given by

$$x(t) = \int_{-\infty}^{\infty} h(\tau)s(t-\tau)d\tau. \quad (1)$$

In [8, 9], the concept of the plenacoustic function has been introduced encapsulating all RIRs of a room for a given source configuration. The PAF, denoted in the following as $p(\boldsymbol{r},t)$, describes the sound field in space depending on both time $t$ and receiver location $\boldsymbol{r} = [r_x, r_y, r_z]^T$. In the simplest case, with the single point source emitting one signal $s(t)$ at fixed position, the PAF is

$$p(\boldsymbol{r},t) = \int_{-\infty}^{\infty} h(\boldsymbol{r},\tau)s(t-\tau)d\tau, \quad (2)$$

where $h(\boldsymbol{r},t)$ is the spatially varying RIR from the source location to the point $\boldsymbol{r}$. Through the LTI system model, the PAF for multiple fixed sound sources consists of a sum of integrals as given in (2). This means, the received sound pressure is a superposition of single source signals, each of them convolved with their specific spatio-temporal RIR. Referring to [8, 9], we consider the PAF, without loss of generality, only for the case where a single source at fixed position emits a Dirac impulse at $t = 0$. With this, the PAF is simplified to the spatio-temporal RIR: $p(\boldsymbol{r},t) = h(\boldsymbol{r},t)$.

For the uniform sampling of a bandlimited RIR or PAF in time, let $T$ denote the sampling interval leading to measurements at equidistant sampling points $t_n = nT$ with $n \in \mathbb{Z}$ being the discrete time variable. According to the Nyquist-Shannon sampling theorem and considering the cutoff frequency $f_c$, the sampling frequency $f_s = 1/T$ has to fulfill the condition $f_s > 2f_c$ in order to avoid aliasing.

### 2.2. Equidistant Sampling in Space

The uniform sampling in space requires a Cartesian grid where the equidistant sampling points $\boldsymbol{r_g} \in \mathcal{G}$ are given by the set

$$\mathcal{G} = \left\{ \boldsymbol{r_g} \mid \boldsymbol{r_g} = \boldsymbol{r}_0 + [g_x\Delta, g_y\Delta, g_z\Delta]^T \right\}, \quad (3)$$

with the grid origin $\boldsymbol{r}_0$ and the discrete grid variables in $\boldsymbol{g} = [g_x, g_y, g_z]^T \in \mathbb{Z}^3$. In order to avoid spatial aliasing, the sampling interval for each spatial dimension $x, y, z$ must follow

$$\Delta < \frac{c_0}{2f_c}, \quad (4)$$

where $c_0$ is the speed of sound [8]. Because of (4), the uniform sampling of the sound field by use of equidistantly spaced microphones often requires an extremely high effort. An array of microphones will most likely never be dense enough to enable measurements without significant problems for very high audio frequencies. For example, the sampling of the PAF with $f_c = 17\,\text{kHz}$ inside a volume of $1\,\text{m}^3$ requires at least $10^6$ spatial measuring points. In order to reduce this infeasible effort, sampling schemes with quasi randomly spaced microphones have been proposed, based on the principle of compressed sensing [12]. However, the maximum admissible audio frequency within larger volumes is still significantly reduced.

A specific problem of microphone arrays is the need for calibration. This includes the compensation of spatio-temporal deviations and the equalization of the frequency responses of the individual microphones. The use of more microphones involves an increasing expense for calibration. In the following section, we propose a dynamic sampling procedure using a moving array with a manageable number of microphones.

## 3. THE PROPOSED DYNAMIC SAMPLING

To determine the sound field within a volume of interest, we consider a scenario in which a single source emits a pre-defined signal and one or more microphones are moved through the volume while their signals are simultaneously recorded together with the microphone-position information.

### 3.1. Model of Virtual Grid in Space

The core of our method is to determine RIRs at equidistant positions by use of dynamic measurements. For that purpose, a virtual sampling grid in space that fulfills (4) is modeled, with integer indices in $\boldsymbol{g}$ spanning the virtual grid coordinate system. The RIRs on that grid are denoted as $h(\boldsymbol{r_g}, t_n)$. The spatial sampling grid is limited to size $X \times Y \times Z$. Overall, the recovery of $h(\boldsymbol{r_g}, t_n)$ inside the finite volume of interest involves $N = XYZ$ RIRs at grid positions $\boldsymbol{g} \in G = \{0, \ldots, X-1\} \times \{0, \ldots, Y-1\} \times \{0, \ldots, Z-1\}$. In the following, the grid RIRs are denoted by $h(\boldsymbol{g}, n)$ using the discrete variables.

### 3.2. The Proposed Method

The RIR at any location within the volume of interest can be computed via interpolation from $h(\boldsymbol{g}, n)$, including the locations on the microphone trajectory. In fact, this interpolation is the key to our method. To keep the following description and analysis simple, a single microphone is considered. The extension to $Q$ microphones is straightforward. Of course, the number of microphones can be traded against the total measurement time, allowing for compromises with a reasonable number of microphones and measurement time. The simplest setup involves only one hand-held microphone which position is tracked. According to (2), each sample $x(n)$ recorded by the dynamic microphone at position $\boldsymbol{r}(n) = [r_x(n), r_y(n), r_z(n)]^T$ contributes an equation of the form

$$x(n) = \sum_{k=0}^{L-1} h(\boldsymbol{r}(n), k)\, s(n-k) + \eta(n) \quad (5)$$

with $\eta(n)$ being the measurement noise. The interpolation of the spatially varying RIR $h(\boldsymbol{r}(n), n)$ from the RIRs at the virtual positions $\boldsymbol{r_g}$ leads to

$$x(n) = \sum_{k=0}^{L-1} \sum_{\boldsymbol{g} \in G} \varphi(\boldsymbol{r}(n), \boldsymbol{r_g})\, h(\boldsymbol{g}, k)\, s(n-k) + \eta(n), \quad (6)$$

where $\varphi(\boldsymbol{r}(n), \boldsymbol{r_g})$ is an interpolation function weighting the sought RIRs on the modeled grid subject to the displacements $\boldsymbol{r}(n) - \boldsymbol{r_g}$. Due to the finite support in space, the virtual grid has to be chosen well below the Nyquist rate (4). Then, an interpolation kernel which is maximally flat in the frequency domain may provide sufficient results. This is shown in the experimental part of this paper. Nevertheless, the interpolation is assumed to be ideal in the following considerations. The sought RIRs on the virtual grid, assumed to be of lengths $L$, are encapsulated in the vector $\boldsymbol{h} \in \mathbb{R}^{NL}$ by the concatenation

$$\boldsymbol{h} = \left[\boldsymbol{h}_1^T, \boldsymbol{h}_2^T, \ldots, \boldsymbol{h}_N^T\right]^T, \quad (7)$$

where $\boldsymbol{h}_u = [h(\boldsymbol{g}_u, 0), \ldots, h(\boldsymbol{g}_u, L-1)]^T$ contains the RIR indexed by $u \in \{1, 2, \ldots, N\}$ on the virtual grid. Using (6) and (7), the system of linear equations

$$\boldsymbol{x} = \boldsymbol{A}\boldsymbol{h} + \boldsymbol{\eta} \quad (8)$$

is set up with measurement vector $\boldsymbol{x} = [x(0), \ldots, x(M-1)]^T$, noise vector $\boldsymbol{\eta} = [\eta(0), \ldots, \eta(M-1)]^T$, and system matrix $\boldsymbol{A} \in \mathbb{R}^{M \times NL}$. The system matrix $\boldsymbol{A}$ possesses the block structure

$$\boldsymbol{A} = [\boldsymbol{\Phi}_1 \boldsymbol{S}, \boldsymbol{\Phi}_2 \boldsymbol{S}, \ldots, \boldsymbol{\Phi}_N \boldsymbol{S}], \quad (9)$$

where $\boldsymbol{\Phi}_u \in \mathbb{R}^{M \times M}$ is a diagonal matrix stacking all $M$ interpolation coefficients for the $u$-th virtual grid RIR,

$$\boldsymbol{\Phi}_u = \mathrm{diag}\{\varphi(\boldsymbol{r}(0), \boldsymbol{r}_{\boldsymbol{g}_u}), \ldots, \varphi(\boldsymbol{r}(M-1), \boldsymbol{r}_{\boldsymbol{g}_u})\}, \quad (10)$$

and $\boldsymbol{S} \in \mathbb{R}^{M \times L}$ is the convolution matrix of the source signal. Ideally, matrix $\boldsymbol{A}$ should have full column rank. Then, the linear system (8) is not underdetermined and its unique least-squares solution yields the estimate of $h(\boldsymbol{g}, n)$. For the underdetermined case, methods of compressed sensing can be used [13, 14], since the sound field holds highly structured sparsity in frequency domain: the spectrum of the PAF in 3D space lives on the 3D surface of a 4D hypercone along the temporal frequency axis [9].

### 3.3. Error Analysis for Spectrally Flat Excitation

Let the excitation signal $s(n)$ be a perfect sequence with period length $L$ and autocorrelation $r_{ss}(m) = \sigma_s^2\,\delta(m \bmod L)$, where $\sigma_s^2$ is the signal power. Thus, for one period in steady state,

$$\boldsymbol{S}\boldsymbol{S}^T = L\sigma_s^2 \boldsymbol{I}_{M \times M}. \quad (11)$$

Let $\boldsymbol{\eta}$ be independent and identically distributed white noise with covariance matrix $\boldsymbol{R}_{\eta\eta} = E\{\boldsymbol{\eta}\boldsymbol{\eta}^T\} = \sigma_\eta^2 \boldsymbol{I}_{M \times M}$. The parameters in $\boldsymbol{h}$ are assumed to have zero mean and variance $\sigma_h^2$, so the autocovariance matrix of $\boldsymbol{h}$ is $\boldsymbol{R}_{hh} = E\{\boldsymbol{h}\boldsymbol{h}^T\} = \sigma_h^2 \boldsymbol{I}_{U \times U}$. Regarding this model, the use of the MMSE estimator for (8) yields the error covariance matrix [15]

$$\boldsymbol{R}_{ee} = E\{[\hat{\boldsymbol{h}} - \boldsymbol{h}][\hat{\boldsymbol{h}} - \boldsymbol{h}]^T\} = \sigma_h^2 \left[\boldsymbol{I}_{U \times U} + \frac{\sigma_h^2}{\sigma_\eta^2} \boldsymbol{A}^T \boldsymbol{A}\right]^{-1}. \quad (12)$$

Using (12) and the relationship

$$\mathrm{tr}\{[\boldsymbol{I}_{S \times S} + \boldsymbol{B}\boldsymbol{C}^T]^{-1}\} = \mathrm{tr}\{[\boldsymbol{I}_{W \times W} + \boldsymbol{C}^T\boldsymbol{B}]^{-1}\} - (W - S) \quad (13)$$

for matrices $\boldsymbol{B}$ and $\boldsymbol{C}$ of size $S \times W$, the MMSE for the model with $M = NL$ can be described as

$$\mathrm{MMSE} = \sigma_h^2 \,\mathrm{tr}\{[\boldsymbol{I}_{M \times M} + \frac{\sigma_h^2}{\sigma_\eta^2}\boldsymbol{A}\boldsymbol{A}^T]^{-1}\}. \quad (14)$$

Following (9), (11), and (14) leads to the estimation error

$$\mathrm{MMSE} = \sigma_h^2 \sum_{n=0}^{M-1} \frac{1}{1 + \frac{\sigma_h^2}{\sigma_\eta^2} L \sigma_s^2 \sum_{u=1}^{N} \varphi^2(\boldsymbol{r}(n), \boldsymbol{r}_{\boldsymbol{g}_u})}. \quad (15)$$

By reference to the expression (15), the estimation error depends on the interpolation coefficients $\varphi(\boldsymbol{r}(n), \boldsymbol{r}_{\boldsymbol{g}_u})$, and, thus, on the trajectory $\boldsymbol{r}(n)$. Given that $\sum_{u=1}^{N} \varphi(\boldsymbol{r}(n), \boldsymbol{r}_{\boldsymbol{g}_u}) = 1$ and assuming nonnegative interpolation coefficients, (15) will be minimal when the coefficients are either zero or one, independent from the actual interpolation accuracy. This means that it would be optimal to sample only on the virtual grid positions. The worst case occurs when the interpolation yields consistently equal coefficients, which corresponds to sampling only in the middle between the grid positions. This worst case can be avoided in practice by considering the grid origin a free parameter and adjusting the grid accordingly.

## 4. TIME-DECOUPLING USING PERFECT SEQUENCES

The periodic excitation by pseudo-random noise $s(n)$ ensuring (11), enables us to decompose the large system (8) with $NL$ unknowns into $L$ smaller systems of linear equations, each having $N$ unknowns. The excitation by $R$ periods of the pseudo-random signal allows for $M = RL$ measurements and leads to the $RL \times L$ convolution matrix

$$\boldsymbol{S}_R = [\boldsymbol{S}^T, \ldots, \boldsymbol{S}^T]^T. \quad (16)$$

Using (8), (9), (11), and $\gamma = L\sigma_s^2$, the measurement process for the noise-free case can be reformulated as

$$\boldsymbol{x} = \sum_{u=1}^{N} \boldsymbol{\Phi}_u \boldsymbol{S}_R \boldsymbol{S}^T \boldsymbol{S} \gamma^{-1} \boldsymbol{h}_u. \quad (17)$$

By defining the modified $RL \times NL$ system matrix

$$\tilde{\boldsymbol{A}} = [\boldsymbol{\Phi}_1 \boldsymbol{S}_R \boldsymbol{S}^T, \boldsymbol{\Phi}_2 \boldsymbol{S}_R \boldsymbol{S}^T, \ldots, \boldsymbol{\Phi}_N \boldsymbol{S}_R \boldsymbol{S}^T] \quad (18)$$

and encapsulating the transformed grid RIRs $\boldsymbol{S}\gamma^{-1}\boldsymbol{h}_u$ into

$$\tilde{\boldsymbol{h}} = [[\boldsymbol{S}\gamma^{-1}\boldsymbol{h}_1]^T, [\boldsymbol{S}\gamma^{-1}\boldsymbol{h}_2]^T, \ldots, [\boldsymbol{S}\gamma^{-1}\boldsymbol{h}_N]^T]^T, \quad (19)$$

the new system of linear equations

$$\boldsymbol{x} = \tilde{\boldsymbol{A}}\tilde{\boldsymbol{h}} + \boldsymbol{\eta} \quad (20)$$

is obtained. Due to (11), the matrix $\tilde{\boldsymbol{A}}$ consists of $R \times N$ blocks of $L \times L$ diagonal matrices which allow to decouple the time dimension by decomposing the large system of size $RL \times NL$ into $L$ smaller problems

$$\boldsymbol{x}_\ell = \tilde{\boldsymbol{A}}_\ell \tilde{\boldsymbol{h}}_\ell + \boldsymbol{\eta}_\ell \quad \ell \in \{1, \ldots L\}, \quad (21)$$

with the sub-vectors $\boldsymbol{x}_\ell$, $\tilde{\boldsymbol{h}}_\ell$, $\boldsymbol{\eta}_\ell$ of length $R$ containing every $L$-th value of the complete vectors, and the sub-system matrix $\tilde{\boldsymbol{A}}_\ell \in \mathbb{R}^{R \times N}$ whose element in row $i$ and column $j$ is given by

$$[\tilde{\boldsymbol{A}}_\ell]_{i,j} = \gamma\,\varphi(\boldsymbol{r}((i-1)L + \ell), \boldsymbol{r}_{\boldsymbol{g}_j}). \quad (22)$$

Due to this re-formulation, the computational demand for reconstructing the sound field is heavily reduced.

## 5. EXPERIMENTS AND RESULTS

For the following experiments, we simulated RIRs and microphone measurements by use of the image source method [16], considering a room of size $5.8\,\mathrm{m} \times 4.15\,\mathrm{m} \times 2.55\,\mathrm{m}$. The reverberation time of the room was chosen as $RT_{60} = 0.3\,\mathrm{s}$. The cutoff frequency of the RIRs was $f_c = 4\,\mathrm{kHz}$. According to the Nyquist-Shannon sampling theorem (4), spatial intervals $\Delta \leq 0.04\,\mathrm{m}$ are required for the virtual grid. The position of the sound source was set to $[1.4, 1.6, 1.0]^T$ in a world coordinate system with unit $1\,\mathrm{m}$. The origin of virtual grid $\mathcal{G}$, as in (3), was set to $\boldsymbol{r}_0 = [2.75, 1.4, 0.8]^T$. Binary MLSs with power $\sigma_s^2 = 1$ and period lengths of $L_p \in \{511, 1023\}$, depending on the lengths of the sought RIRs, were used as excitation signals in the experiments. The measurements were simulated in a steady-state room, assuming periodic excitation starting at $n = -L_p$. We sampled the PAF on a plane at height $0.8\,\mathrm{m}$ by setting $Z = 1$ in $G$. The 3D problem can be seen as a stack of multiple plane grids.

As evaluation criterion for the quality of the PAF, we use the *mean normalized system misalignment* [17, 18]

$$\mathrm{MNSM} = \frac{1}{N}\sum_{u=1}^{N} \frac{\|\boldsymbol{h}_u - \hat{\boldsymbol{h}}_u\|_{\ell_2}^2}{\|\boldsymbol{h}_u\|_{\ell_2}^2}. \quad (23)$$

**Table 1**. MNSM [dB] of the RIRs on a $5 \times 5$ grid, depending on levels of measurement noise. Results for the static method with 25 microphones and the new proposed dynamic method using $Q \in \{25, 20, 15, 10, 5\}$ microphones. Measurements are performed exactly on grid positions.

| Sampling | SNR [dB] | | | | | | |
|---|---|---|---|---|---|---|---|
| | 10 | 20 | 30 | 40 | 50 | 60 | 70 |
| Static | 0.43 | -9.55 | -19.53 | -29.57 | -39.45 | -49.45 | -59.51 |
| Dyn-25 | 0.49 | -9.60 | -19.61 | -29.49 | -39.47 | -49.58 | -59.52 |
| Dyn-20 | 1.71 | -8.28 | -18.28 | -28.37 | -38.29 | -48.40 | -58.26 |
| Dyn-15 | 3.10 | -6.94 | -16.98 | -26.98 | -37.00 | -46.97 | -56.97 |
| Dyn-10 | 4.72 | -5.31 | -15.26 | -25.31 | -35.27 | -45.33 | -55.27 |
| Dyn-5 | 7.51 | -2.49 | -12.59 | -22.47 | -32.47 | -42.50 | -52.58 |

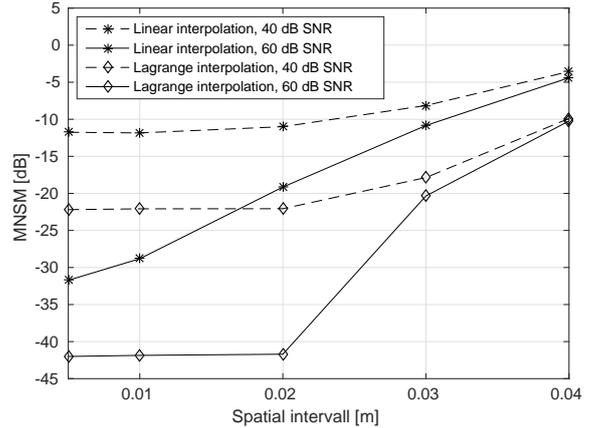

**Fig. 1**. Recovery results for a Lissajous trajectory, depending on the spatial sampling interval of the virtual grid and interpolation method.

with $\boldsymbol{h}_u \in \mathbb{R}^L$ containing the true RIR and $\hat{\boldsymbol{h}}_u \in \mathbb{R}^L$ being the reconstructed RIR at grid index $u$.

The first experiment is devoted to a proof of concept of the general validity of the new approach. For this, we compare the best possible dynamic sampling with the conventional static technique. That is achieved by constraining the sampling positions of the dynamic procedure to the points of the spatial grid. This avoids all interpolation errors and additionally leads to optimal interpolation coefficients according to (15). The spatial grid was modeled with sampling interval $\Delta = 0.02$ m and $X = Y = 5$. The lengths of the RIRs were set to $L = 500$. An MLS with $R = 10$ periods and period length $L_p = 511$ was used. For the static method, we measured the 25 RIRs on the spatial grid by correlating each measurement signal with one period of the MLS and averaging over all $R$ periods [3, 4]. For estimation of the baseline of the recovery quality, different noise conditions were considered. The MNSMs for different signal-to-noise ratios SNR $= \sigma_s^2/\sigma_\eta^2$ can be seen in the first line of Table 1. For the dynamic method, five different setups were considered. The first one used also $Q_{\max} = 25$ microphones. The dynamic sampling was achieved by random rotations of the microphone array around its center. After taking a sample, the array was rotated by a multiple of $\pi/2$. This method fulfills the requirement $\boldsymbol{r}_q(n) \in \mathcal{G}$ for all $n$ and all microphones $q$. The other setups used a smaller number $Q \in \{20, 15, 10, 5\}$ of microphones which allowed additional translations with a multiple of $\Delta$. For the dynamic recovery, we tested both, the least-squares solution of the large system (8), and the least-squares solutions of the 511 time-decoupled systems (21). Both strategies led to the same recovery results, which are also given in Table 1. The comparison with the first line shows that the dynamic sampling with 25 microphones is as good as the static one for all noise levels. The reduction of the number of microphones leads to a smaller number of measurements and therefore a smaller number of equations describing the sought RIRs. The reduced quality of reconstruction follows closely $\Delta_{\text{MNSM}} = 10 \cdot \log_{10} (Q_{\max}/Q)$ for all noise levels.

In a further experiment, we tested the dynamic recovery technique for spatial grids with different sampling intervals. Each of the virtual grids involved 400 RIRs with $X = Y = 20$. The lengths of the RIRs were set to $L = 1000$. An MLS with $L_p = 1023$ was used for excitation. The dynamic measurements were taken by only one moving microphone over $R = 1000$ periods. Solving the system (8) for recovery would involve $4 \cdot 10^5$ unknowns in this case. Hence, we decoupled the time dimension and solved the linear systems (21) for least-squares, each comprising only $4 \cdot 10^2$ unknowns. For trajectories with uniformly distributed measurement positions on the grid, we observed that the accuracy of the sound field recovery decreases on outer grid positions. Hence, we tested Lissajous trajectories, which provide a high density of sampling points in curves at the boundary of the grid. The results for Lissajous trajectory with frequency ratio of $17/16$ are shown in Fig. 1. We tested both the linear interpolation and the Lagrange interpolation for reconstruction. For the uniform grid in space, the Lagrange interpolator is equivalent to a maximally flat finite impulse response filter, as both obtain the same coefficients [19]. Here, the maximum degree of the interpolation polynomial was limited to 19, depending on the measurement position.

In the case of sampling at the Nyquist rate ($\Delta = 0.04$ m), both interpolation methods lead to almost the same recovery quality. With spatial oversampling, the performance increases in all cases. However, the Lagrange interpolation gives a clearly better reconstruction than the linear one. Additionally, its performance shows a plateau for more than a twofold spatial oversampling ($\Delta < 0.02$ m).

## 6. CONCLUSION

In this paper, we proposed a new method for the measurement of sound fields using moving microphones. With known trajectories and excitation signal, a system of linear equations has been derived, which leads to the estimation of spatially dependent RIRs on an equidistant virtual grid. In order to reduce the computational complexity perfect sequences has been used. The method has been tested using two different interpolators and spatial intervals. A potential application of the proposed method is the high-precision measurement of sound fields using hand-held microphones whose positions are continuously tracked, e.g., using gyroscopes. Further extensions based on the theory of compressed sensing for trajectories that lead to ill-conditioned linear systems are currently under investigation.